\documentstyle[multicol,aps,prb,epsf]{revtex}

\newcommand{\Vec}[1]{\mbox{\boldmath$#1$}}

\begin{document}

\draft

\title{Hofstadter butterfly and 
integer quantum Hall effect in three dimensions}
\author{M. Koshino, H. Aoki, K. Kuroki} 
\address{Department of Physics, University of Tokyo, Hongo, Tokyo
113-0033, Japan}
\author{S. Kagoshima}
\address{Department of Basic Science, University of Tokyo, Komaba, Tokyo 
153-8902, Japan}
\author{T. Osada}
\address{Institute for Solid State Physics, University of Tokyo,
Kashiwa, Chiba 277-8581, Japan}

\date{\today}

\maketitle

\begin{abstract}
For a three-dimensional lattice in magnetic fields 
we have shown that 
the hopping along the third direction, which normally tends to 
smear out the Landau quantization gaps, 
can rather give rise to a fractal energy spectram akin to 
Hofstadter's butterfly when a 
criterion, found here by mapping the problem to two dimensions, 
is fulfilled by anisotropic (quasi-one-dimensional) systems.   
In 3D the angle of the magnetic field plays the role of the field 
intensity in 2D, so that the butterfly can occur in much smaller fields.  
The mapping also enables us to calculate the Hall 
conductivity, in terms of the 
topological invariant in the Kohmoto-Halperin-Wu's
formula, where each of $\sigma_{xy}, \sigma_{zx}$ is found to be quantized.
\end{abstract}

\begin{multicols}{2}
\narrowtext

Among the effects of magnetic fields on electronic states, 
one of the most bizzare is Hofstadter's butterfly.  
Namely, when a two-dimensional (2D) 
periodic system is put into a magnetic field, 
the gap not only appears between the 
Landau levels, but a series of gaps appear in a selfsimilar fashion 
as shown by Hofstadter\cite{Hofs}. 
The butterfly refers to an energy spectrum 
against the magnetic flux, $\phi$, penetrating a unit cell in units of 
the flux quantum $\phi_0=h/e$.  Usually the butterfly is conceived to be a 
phenomenon specific to 2D.  

Here we raise a question: can we have something like Hofstadter's 
butterfly in spite of, 
or even because of, a three-dimensionality (3D)?  
This may at first seem quite unlikely, since the usual derivation 
of the butterfly relies on the two-dimensionality of the system, so that 
a motion along the third direction ($z$) 
should tend to wash out the butterfly gaps as well as Landau level gaps.  
Several authors have extended
Hofstadter's problem to 3D in the last decade \cite{Hase,Kuns}, 
and subbands are indeed shown to overlap or touch with each other.  
Here we systematically look for the possibility of butterflies 
(i.e., recursive and fractal gaps) in 3D.  

If we do have a butterfly, then we can proceed to question 
how the integer quantum Hall effect should look like on the butterfly.  
If one examines a theoretical reasoning from which the 
quantization in the Hall conductivity is deduced in the 
usual quantum Hall system, the essential ingredient 
is the presence of a gap in the energy spectrum.  
This was indicated in a gauge argument by Laughlin\cite{Laughlin}, 
and elaborated by 
Thouless, Kohmoto, Nightingale, and den Njis \cite{Thou} 
when a periodic potential exists.  
There the quantized Hall conductivity 
when the Fermi energy, $E_F$, lies in a butterfly gap 
is identified to be a topological invariant 
characterizing the position of gaps.

For 3D Kohmoto, Halperin, and Wu have shown, 
following the line of the 2D work, 
that {\it if} there is an energy gap in a 3D system, then an integer 
quantum Hall effect should result 
as long as $E_F$ lies within a gap\cite{Halp,Kohm}. 
Montambaux and Kohmoto have actually calculated the Hall conductivity 
in a case where a third-direction hopping opens some gaps.\cite{Mont} 
If butterflies are realized in 3D, 
we can move on to the systematics of the quantum Hall numbers.  

In the present paper we point out that, first, 
an analog of Hofstadter's butterfly does indeed exist specific to 3D 
under certain criterion that is fulfilled by anisotropic (quasi-1D) 
tight-binding lattices in 3D.  In this case 
the butterfly plot refers to energy versus {\it angle} of the magnetic 
field.  This is obtained by mapping the 3D system to a 2D system.  
Remarkablly the mapping dictates 
that the ratio of the magnetic fluxes penetrating 
two facets of the unit cell plays the role of the magnetic flux 
in 2D, so that the field intensity 
$B$ does not have to be strong to realize the butterfly.  

More importantly the mapping enables us to systematically calculate the Hall 
conductivity for the 3D butterfly via identifying the 
topological invariant in the Kohmoto-Halperin-Wu's
formula.  We have found that {\it each of $\sigma_{xy}, 
\sigma_{zx}$ is quantized} in 3D.  

Our model is non-interacting tight-binding electrons in a
uniform magnetic field $\Vec{B}$ described by the Hamiltonian,
\begin{equation}
{\cal H} = 
\sum_{\langle i,j\rangle} (t_{ij} e^{i\theta_{ij}} c^{\dagger}_i c_j +
{\rm h.c.}), \label{Schr}
\end{equation}
in standard notations, where 
the summation is taken over nearest-neighbor sites 
with $t_{ij}=t_x, t_y, t_z$ along $x,y,z$, respectively, 
$\theta_{ij} = (e/\hbar)\int_i^j\Vec{A}\cdot d\Vec{l}$ is the Peierls phase. 
We first recapitulate the 2D case, because a 
key in this work is a correspondence between 2D and 3D. 
In 2D with the Landau gauge $ \Vec{A} = (0,Bx)$, $y$
is a cyclic coordinate, so that the wave function becomes 
$\psi_{lm} = e^{i \nu_y m} F_l$, 
where $(l,m)$ labels $(x,y)$ coordinates, 
and $\nu_y$ is the Bloch wave number along $y$. The
Schr\"{o}dinger equation then takes a form of Harper's equation, 
\begin{equation}
-t_x(F_{l-1} \,+\, F_{l+1}) \,-\, 2t_y \cos(2\pi \phi l +\nu_y )F_{l}= E F_{l}\label{Harp2D},
\end{equation}
where $\phi=Bab/\phi_0$ is the number of flux
quanta penetrating a unit cell $=a\times b$.
While the energy spectrum becomes 
a butterfly for the ordinary isotropic case, $t_x=t_y$, 
the gaps are rapidly smeared out as the anisotropy is 
introduced, $t_y/t_x \rightarrow 0$, since the potential term (the 
cosine function above) weakens.  
Since $t_x$ and $t_y$ appear on an equal footing, 
the condition for the appearance of the butterfly in 2D is 
$t_x \approx t_y$.

\begin{figure}
\begin{center}
  \leavevmode\epsfxsize=80mm \epsfbox{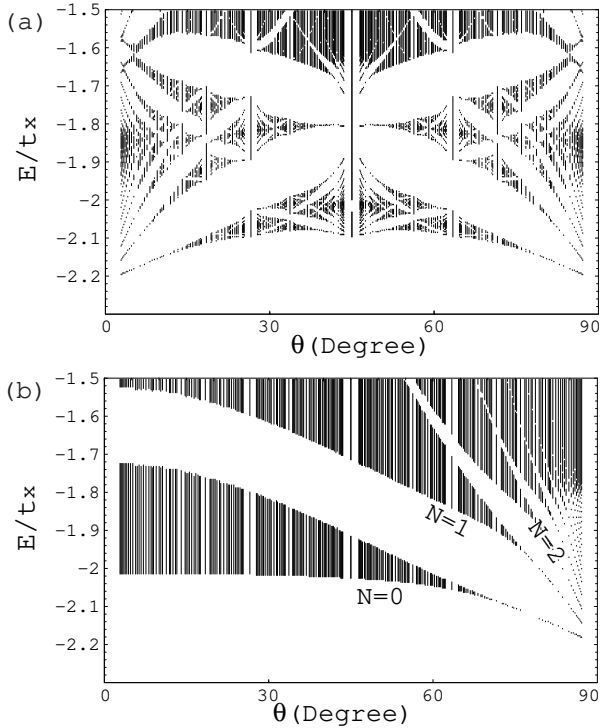}
\end{center}
\caption{The energy spectrum of a 3D system 
with $t_x:t_y:t_z = 1:0.1:0.1$ (a) or a 2D system 
with $t_x:t_y:t_z = 1:0.1:0$ (b) plotted
  against the angle $\theta$ in a magnetic field $(\phi_y, \phi_z) =
  0.2(\sin\theta, \cos\theta)$.}
\label{btfl3D}
\end{figure}

Harper's equation in 3D can be derived in a similar 
way.  For simplicity we consider a simple cubic lattice in a
magnetic field $\Vec{B}=(0,B\sin\theta, B\cos\theta)$ 
assumed to lie in the $yz$ plane. The vector potential is then 
$\Vec{A}=(0,Bx\cos\theta,-Bx\sin\theta)$, so that 
$y,z$ are cyclic and the wave function becomes 
$\psi_{lmn} = e^{i \nu_{y} m + i \nu_{z} n} F_l$,
where $(l,m,n)$ labels $(x,y,z)$.
The Schr\"{o}dinger equation is 
\begin{eqnarray}
  -t_x(F_{l-1} \,+\, F_{l+1}) \,-\,  [2t_y \cos(2\pi \phi_{z}l
  +\nu_{y}) \nonumber \\ + 2t_z \cos(-2\pi \phi_{y}l +\nu_{z})] F_{l} = E
  F_{l}\label{Harp3D},
\end{eqnarray}
where two periodic potentials are now superposed.  
Here $\phi_{y} (\phi_{z})$ 
is the number of flux quanta penetrating the side of a 
unit cell ($=a\times b \times c$) normal to $y (z)$ 
(inset of Fig.\ref{btflHall3D}(a)).

Although the spectrum in 3D does not in general have many gaps 
(aside from the trivial Bragg-reflection gaps), 
we first notice that a butterfly-like structure does emerge 
for certain choices of $(t_x,t_y,t_z)$, as 
typically displayed in \ref{btfl3D}(a) for 
$(t_x,t_y,t_z)=(1, 0.1, 0.1)$, a quasi-1D system. 
The spectrum is plotted against the 
angle $\theta$ of a magnetic field $(\phi_y,\phi_z) =
0.2(\sin\theta,\cos\theta)$ with $b=c$ assumed here.  
A structure akin to the 2D butterfly is seen in the bottom (or 
at the top) of the whole spectrum. 
One might consider this as a 2D butterfly surviving the third-direction 
hopping, but this is wrong as is evident from Fig.1(b), 
where we turn off $t_z$ to find that 
the spectrum coalesces to a series of broadened Landau levels.   
So we are talking about the butterfly {\it specific} to 3D 
rather than a remnant of a 2D butterfly.  

We first explore the mechanism why the butterfly appears in 3D.  
The doubly periodic potential in the 3D Harper equation comprises 
$V^{(1)}(l) \propto t_y \cos(2\pi \phi_{z}l +\nu_{y}), 
V^{(2)}(l) \propto t_z \cos(-2\pi \phi_{y}l +\nu_{z})$.  
We assume that their periods, $1/\phi_z$ and $1/\phi_y$, 
are much greater than the lattice constant ($\phi_z,\phi_y \ll 1$).  
We also assume that 
\begin{equation}
t_y\phi_z \gg t_z\phi_y,
\label{Assump2}
\end{equation}
which amounts to an assumption that the
local peaks and dips of the total potential
$V^{(1)}+V^{(2)}$ is primarily that of $V^{(1)}$. 

One can then regard the potential minima of $V^{(1)}$ as 
`sites', which we call `wells' to distinguish from the original
sites. The wells are separated by $1/\phi_z$ and feel the 
slowly-varying $V^{(2)}$. Since each well contains many original
sites due to the first assumption, we can talk about bound states for
the well in the effective-mass approach. If wells are deep 
enough, several bound states appear and each state forms a 
tight-binding band (i.e., Landau band), 
and the equation (\ref{Harp3D}) reduces to
\begin{eqnarray}
  -t'(J_{l'-1} \,+\, J_{l'+1}) \,-\, 2t_z \cos[-2\pi (\phi_y /
  \phi_z)l' \nonumber \\ + (\phi_y / \phi_z)\nu_{y} + \nu_z] J_{l'} =
  E J_{l'}\label{HarpEff}.
\end{eqnarray}
Here $t'$ is a transfer integral between neighboring 
wells labelled by $l'$, $J_{l'}$ the `effective-mass' 
wave function, and the cosine term is 
the value of $V^{(2)}$ at each minimum of $V^{(1)}$.  
The reduced equation has exactly the same form as that of the 2D system,
eqn.(\ref{Harp2D}) if we translate 
\begin{equation}
{\rm 3D:}(t_x,t_y,t_z,\phi_y,\phi_z) \longrightarrow 
{\rm 2D:}(t',t_z,\phi_y / \phi_z).
\label{2D3Dcorresp}
\end{equation}
Since the butterfly is a hallmark of an isotropic 2D case, 
one can predict that the subband in 3D for which $t' \approx t_z$
should exhibit a butterfly.  

We can estimate $t'$ by applying the effective-mass approximation
to Harper's equation (\ref{Harp3D}). 
We first convert the 
equation (when there is $V^{(1)}$ alone) to a differential 
equation for a continuous variable $\tilde{l}\equiv 2\pi \phi_z l$, 
which turns out to contain a combination $t_y/\phi_z^2$ only 
(with $t_x=1$, a unit of energy).  Since $t'$ 
is a matrix element of $V^{(1)}\propto t_y$, we have a simple scaling law, 
\begin{equation}
t' = 2t_y \, \, f \left( \frac{t_y}{\phi_z^2} \right).
\end{equation}
Note that the value of $t'$ differs from one tight-binding band to another, 
where middle bands, with weaker binding, have larger $t'$.  

\begin{figure}
\begin{center}
\leavevmode\epsfxsize=80mm \epsfbox{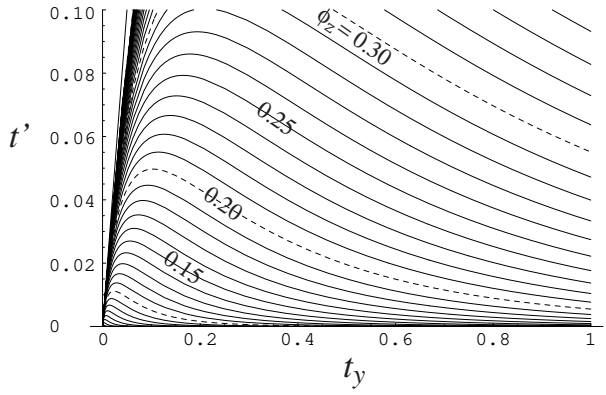}
\end{center}
\caption{
  The effective transfer $t'(t_y,\phi_z)$ for the lowest
  Landau band in units where $t_x=1$. 
  By plugging $t'\simeq t_z$, the plot serves to identify 
  appropriate values of $\phi_z$ to 
  realize the butterfly for given $(t_y,t_z)$.
}
\label{effectiveT}
\end{figure}

\begin{figure}
\begin{center}
  \leavevmode\epsfxsize=80mm \epsfbox{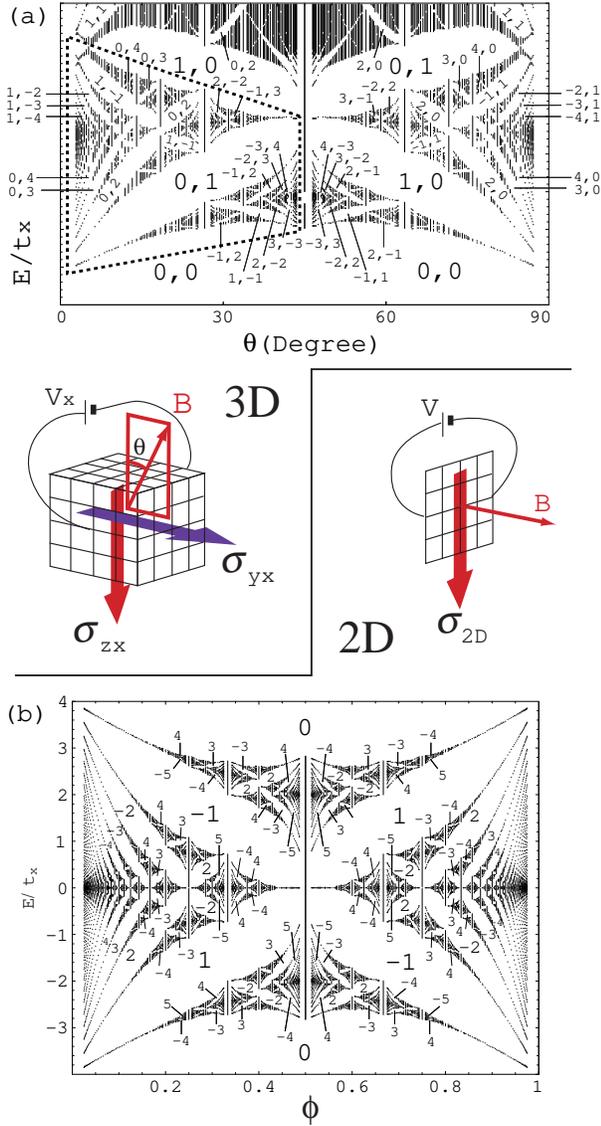}
\end{center}
\caption{
  (a) The conductivity $(\sigma_{xy},\sigma_{zx})
  = -e^2/h (\mu_3,\mu_2)$ is plotted on a 3D butterfly, where we
  display the topological invariants $(\mu_3,\mu_2)$ for each gap.
  (b) The corresponding plot for $\sigma _{2D} = -(e^2/h)t$ on the 2D
  butterfly. The area enclosed by a dashed line in (a) corresponds to
  the 2D butterfly. $\mu_2$ in (a) corresponds to $t$ in (b), while
  $\mu_3$ in (a) to $s$ in Eq.(\ref{CorresInteg})}
\label{btflHall3D}
\end{figure}

We have then numerically 
calculated $t'(t_y, \phi_z)$ for the lowest band.  
With the scaling, the $t_y$-dependence of $t'$ for a particular $\phi_z$ 
provides the whole dependence, shown in Fig.\ref{effectiveT}.  
If we plug in the condition for the butterfly, $t_z \simeq t'$, 
the plot indicates how to adjust $\phi_z$ to have a 
butterfly for given $(t_y,t_z)$.  One can immediately 
find that the butterfly is restricted to the case with 
$t_y, t_z \ll 1(=t_x)$, i.e., quasi-1D systems.  
This is because $\phi_z$ becomes too large to satisfy $\phi_z \ll 1$ 
in the most region of $t_y \approx 1$ or in the
region $t_z \approx 1$ (out of the plot).  

This plot shows that the above example, 
$(t_x,t_y,t_z) = (1,0.1,0.1)$, is indeed a right choice, 
for which we can show that 
$t'=0.05\approx t_z (= 0.1)$ for the lowest subband.  
Incidentally, the butterfly is symmetric in this example, which is an accident 
for $t_y=t_z$: in Harper's equation $V^{(1)}$ and $V^{(2)}$ exchange
roles at $\theta=45^\circ$ for $t_y = t_z$.  Around $45^\circ$, 
or more generally around $t_y\phi_z \approx t_z\phi_y$, 
the above argument breaks down but a clear structure remains.  
We can explain this as follows.  
When $t_y\phi_z \approx t_z\phi_y$, $V^{(1)}+V^{(2)}$ 
exhibits a beat so that the barrier height separating the wells 
varies from place to place, which implies that 
$t'$ varies from place to place. 
However, a change in the wall height ($t_y$) changes $t'$ only slightly, 
since $t'$ has a broad peak in Fig.\ref{effectiveT}. 

Thus in 3D the effective flux in eqn.(\ref{2D3Dcorresp}) 
is a ratio $\phi_y/\phi_z$.  
In 2D by contrast, a butterfly requires $\phi \sim O(1)$, i.e., 
$B$ has to be impossibly large ($\sim 10^5$T for $a=2$\AA).  
Would this render the 3D butterfly experimentally feasible?\cite{AMRO}  
In principle, Fig.\ref{effectiveT} shows that
there exists appropriate $(t_y,t_z)$ no matter how small $\phi_z$ may be.  
In practice, $(t_y,t_z)$ become smaller as $\phi_z$ decreases, 
and the energy scale (width of the Landau band $4(t'+t_z)$ 
with $t'\approx t_z$) shrinks for smaller $\phi_z$, so that it will be hard to
resolve the butterfly structure.  
For typical quasi-1D organic conductors such as (TMTSF)$_2$X 
we have $t_x:t_y:t_z=1:0.1:0.01$ with $a, b, c \sim 10 {\rm \AA}$, 
and we can estimate the required $\phi_z \sim 0.1$, 
which corresponds to $B \sim 400$T.  
It is still huge, but much smaller than the value required for 2D and 
around the border of experimental feasibility.\cite{commentTMTSM}

While we have discussed appropriate values for given $(t_y,t_z)$, 
are there restrictions on $(t_y,t_z)$ to have butterflies? 
Binding of a well must be so strong that the transfer to second
neighbors is negligible.  If one approximates a
well in $V^{(1)}$ with a harmonic potential, the quantized energies are 
$(n+\frac{1}{2})\hbar\omega$.  Then 
the $n$-th state is strongly bound to each well when this energy is 
smaller than $4t_y$, the depth of a well.  Hence 
$t_y$ should not be too small ($\sqrt{t_y}>\phi_z$), otherwise 
we have trivial Bragg-reflection gaps only. 
Also, the residual potential $V^{(2)}$ whose amplitude is $2t_z$ 
must be weaker than $\hbar\omega$ (i.e., $t_z<\phi_z\sqrt{t_y}$) so that
different Landau bands are not mixed.  
All the conditions (those discussed in this paragraph as well 
as $\phi_z,\phi_y \ll 1, t_y\phi_z \gg t_z\phi_y$) 
can be interpreted in the semiclassical quantization involving 
the cross sections of equipotential surfaces, but the essential 
hopping ($t'$) between adjacent cross-sectional orbits is outside 
the scope of the semiclassical picture.

Now we come to our goal of calculating the Hall conductivity 
when the Fermi level lies in each gap of the 3D butterfly.   
The mapping does indeed enables us to accomplish this through identifying 
the topological invariants in the general formula for the
Hall conductivity for 3D Bloch electrons by Kohmoto-Halperin-Wu.  
In the formula the Hall conductivity tensor is expressed as
\begin{equation}
\sigma_{ij}= - \frac{e^2}{2\pi h}\sum_k \epsilon_{ijk}G_k
\end{equation}
when Fermi energy is in a gap.  
Here $\epsilon_{ijk}$ is a unit antisymmetric tensor, 
$\Vec{G} = \mu_1\Vec{a^*} + \mu_2\Vec{b^*} + \mu_3\Vec{c^*}$ 
with $\Vec{a^*},\Vec{b^*},\Vec{c^*}$ being the primitive reciprocal
lattice vectors, and $\mu_1,\mu_2,\mu_3$ are topological invariants specifying 
each gap (i.e., remain constant when we change the direction of 
${\boldmath B}$ in the present context).  
For an orthogonal lattice we have simply 
$\sigma_{yz} = -\frac{e^2}{h}\frac{\mu_1}{a}, \, 
\sigma_{zx} = -\frac{e^2}{h}\frac{\mu_2}{b}, \,
\sigma_{xy} = -\frac{e^2}{h}\frac{\mu_3}{c}$.  

So we have only to determine invariant integers, which are subject to 
a Diophantine equation, 
\begin{equation}
\frac{r}{Q} = \lambda + 
\frac{P}{Q}n_x \mu_1 + \frac{P}{Q}n_y \mu_2 + \frac{P}{Q}n_z \mu_3,
\label{Dio}
\end{equation}
where we have assumed a rational magnetic flux, 
$(\phi_x,\phi_y,\phi_z)=\frac{P}{Q}(n_x,n_y,n_z)$ ($P,Q$: integers, 
$n_x$ etc have no common divisors), 
$r$ the number of occupied bands, and $\lambda$ another topological invariant.
Although the solution of the Diophantine equation is not unique, 
Thouless {\it et al.}\cite{Thou} argued for 2D that there is a restriction
on the integers that decides the solution uniquely. 
Kohmoto {\it et al.} \cite{Kohm} 
conjecture the uniqueness of the solution in 3D 
in analogy with the 2D case, where the restriction 
for 3D reads $|\mu_1 n_x + \mu_2 n_y + \mu_3 n_z| < Q/2$.

We can then calculate the Hall conductivity for the 3D butterfly. 
We assume 
$(\phi_x,\phi_y,\phi_z) = P/Q(0, n_y,n_z)$ and $t_y\phi_z \gg t_z\phi_y$, 
for which the effective flux in eq.(\ref{HarpEff}) is
$\phi=\phi_y/\phi_z=n_y/n_z$.  Hence each Landau band should split
into $n_z$ butterfly subbands. Let us consider the situation where 
$E_F$ lies just above the $m$th subband in the $l$th Landau
band from the bottom, i.e., $(l n_z + m)$ subbands altogether.  
Each subband is shown to comprise $P$
bands, so that the gap has an index $r = (l n_z + m)P$.  
Substituting this in the Diophantine eq.(\ref{Dio}), we have 
$(l n_z + m)P = Q \lambda + P n_y \mu_2 + P n_z \mu_3$.
Since $P$ and $Q$ have no common divisors, $s$ must be a multiple of
$P$, and with the above restriction 
one has $\lambda=0$, and we end up with 
$l n_z + m =  n_y \mu_2 + n_z \mu_3$ 
for the 3D butterfly in the lower 
half of the entire band. Thus we can
determine $\mu_2,\mu_3$ for an arbitrary gap in the 3D
butterfly as explicitly displayed in Fig.\ref{btflHall3D}(a).

If we compare with a corresponding plot for 2D in Figure \ref{btflHall3D}(b), 
we recognize a beautiful consequence of the 2D-3D mapping established here 
as a unsuspected one-to-one correspondence between 
the Hall conductivities on 2D and 3D butterflies {\it as a whole} 
(i.e., for a set of topological invariants attached to the recursive gaps). 
Namely, the Hall conductivity in 2D\cite{Thou} is given by
$\sigma_{\rm 2D} = -\frac{e^2}{h}t$,
where $t$ is an integer in a 2D Diophantine equation
$r =  qs + pt$.  If we compare this with 
the 3D Diophantine equation, $m =  n_z (\mu_3 - l) + n_y \mu_2$, 
the mapping dictates a correspondence
$n_y \leftrightarrow p, \,n_z \leftrightarrow q, \,
m \leftrightarrow r$, 
so that the invariant integers should translate as
\begin{equation}
\mu_3-l \longleftrightarrow s, \,
\mu_2   \longleftrightarrow t. \label{CorresInteg}
\end{equation}
This implies that $\sigma_{zx}$ in 3D plays the role of $\sigma_{\rm 2D}$.

Future problems are to extend the effective theory to the
the arbitrary orientation of $\Vec{B}$, for which Harper's equation
in the arbitrary $\Vec{B}$ has been discussed in an existing
literature \cite{Hase,Kuns}. 
We wish to thank Mahito Kohmoto for discussions.

\end{multicols}

\begin{thebibliography}{99}
\bibitem{commentTMTSM}
Electron-electron interaction may not be 
negligible, which, e.g., gives rise to SDW in TMTSF compounds 
in strong magnetic fields.
\end{thebibliography}
\end{document}